\documentclass[preprint,preprintnumbers,amsmath,amssymb,prb]{revtex4}

\usepackage{graphicx}
\usepackage{dcolumn}
\usepackage{bm}
\usepackage{amssymb}
\usepackage{latexsym}

\begin{document}
\title{Phase Behavior of Short Range Square Well Model}
\author{D. L. Pagan and J. D. Gunton
      \\    \emph{\small{Department of Physics, Lehigh University, Bethlehem, P.A.
       18015}}}
\date{\today}

\begin{abstract}
    \noindent Various Monte Carlo techniques are used to determine the
    complete phase diagrams of the square well model for the attractive
    ranges $\lambda = 1.15$ and $\lambda = 1.25$.  The results for the latter case
    are in agreement with earlier Monte Carlo simulations for the fluid-fluid coexistence curve and yield
    new results for the liquidus-solidus lines. Our results for $\lambda = 1.15$ are new. We find that the fluid-fluid
    critical point is metastable for both cases, with the case $\lambda = 1.25$ being just below the threshold value for
    metastability. We compare our results with prior studies and with experimental results for the gamma-II crystallin.
\end{abstract}

\maketitle

\section{Introduction}
    It has been known since the the work of Gast \emph{et
    al.}~\cite{kn:gast}, and subsequently confirmed by
    others~\cite{kn:leker,kn:experiment1}, that the phase behavior of colloidal particles
    depends sensitively on the range of attraction between them.
    For sufficiently short range attractive interactions, the phase diagram exhibits a solid-fluid coexistence curve
that is subtended by a metastable fluid-fluid coexistence curve.
Such phase behavior is also typical of certain globular proteins
in solution~\cite{kn:experiment2}. This has led  scientists to
model globular proteins in solution using the ideas developed in
colloid science.
    Several models with short range attractive interactions ~\cite{kn:yukawa,kn:tenwolde,kn:oxtoby,kn:benedek} have been used to
    calculate the phase diagrams of various proteins using computer simulation.
    As well, many of these models have been shown
    to obey a kind of van der Waal's extended states behavior~\cite{kn:extended}. These isotropic
    short-range models have also been used to characterize~\cite{kn:tenwolde,kn:oxtoby} the
     nucleation rates for globular proteins.
    The simplest of these models is the square well potential,
    given by
    \begin{equation}
        \large{V(r)} = \left\{ \begin{array}{ccc}
                        \infty, &\mbox{$r < \sigma$}  \\
                        -\epsilon, &\mbox{$\sigma \le r <
                        \lambda\sigma$}\\
                        0, &\mbox{$r \ge \lambda\sigma$}.
                   \end{array} \right.
    \end{equation}

    \noindent The fluid-fluid coexistence curve for values of $\lambda \ge
    1.25$ is known~\cite{kn:vega}. However, to date, no direct simulation results
    of the complete phase diagram
    are available for $\lambda = 1.25$ or for smaller ranges away from the adhesive-sphere limit. This choice of
     $\lambda = 1.25$    for the
the range of
    interaction is believed to be close to the threshold value below which the
    square well model becomes metastable. Indeed, theoretical and semi-analytical treatments confirm this
    hypothesis~\cite{kn:daanoun,kn:benedek}, although the threshold values for other short-range models is smaller.
    Also, the phase behavior at this value for the square well model
    has been compared to that of the gamma crystallin proteins~\cite{kn:crystallins}, mutants of which
    have been linked to genetic cataracts~\cite{kn:mutation2}. The square well model has also been used to determine nucleation rates
    at several ranges of $\lambda$ to better understand protein crystallization~\cite{kn:dixit} and is also of
    interest for comparing results with theoretical approaches modelling phase behavior~\cite{kn:stell}, as well as for comparison
with experimental studies. It is important,
    therefore, that these phase diagrams be accurately calculated, both at and below $\lambda = 1.25$.  In this paper, we present such results using standard
    Monte Carlo and parallel tempering techniques.

\section{Computational Methods and Details}
\subsection{Solid-fluid coexistence}

    Kofke~\cite{kn:kofke} showed that the solid-fluid
    coexistence curve can be obtained by solving the first-order
    Clausius-Clapeyron equation, given by~\cite{kn:note}

    \begin{equation}
        \frac{dP}{d\beta} = -\frac{\triangle h}{\beta \triangle v}
    \end{equation}

    \noindent where P is the pressure, $\beta = 1/T$ is the
    inverse temperature, h is the molar enthalpy, and v is the
    molar volume, respectively. The method, known as the
    Gibbs-Duhem method, requires that two isobaric-isothermal (NPT) Monte Carlo simulations be
    carried out in parallel so that information can be used to calculate the next state and, thus, points along the
    whole coexistence curve can be obtained. This method has been
    employed for solid-fluid coexistence for a variety of models~\cite{kn:kofke,kn:noro,kn:yukawa}.
    One \emph{caveat} is that an initial coexistence point must
    first be known in order to integrate the Clausius-Clapeyron
    equation. To do this, the equation of state along an isotherm
    is obtained using NPT simulations. Once the equation of state
    is known, we fit the fluid-solid lines with an equation of the
    form

    \begin{equation}
        \beta P = \frac{\rho}{1-a\rho} +
        b(\frac{\rho}{1-a\rho})^{2}
        + c(\frac{\rho}{1-a\rho})^{3}
    \end{equation}

    \noindent where $\rho$ is the number density. The solid line is fit to a $2^{nd}$ order
    polynomial equation of the form $a\rho^{2} + b\rho +c$.
    Integrating these two equations yields the chemical potentials
    for the liquid and solid, as shown in equations (4) and (5), respectively.

    \begin{equation}
        \beta \mu_{l} = \ln(\frac{\rho \Lambda^{2}}{1 - a\rho}) +
        \frac{b/a - c/a^{2} + 1}{1 - a\rho} + \frac{c/2a^{2} +
        b\rho}{(1-a\rho)^{2}}+ \frac{c\rho^{2}}{(1 - a\rho^{3})} -
        (b/a - c/2a^{2} + 1)
    \end{equation}

    \begin{equation}
      \beta \mu_{s} = 2a\rho + b(\ln \rho + 1) - (a\rho^{*} + b\ln \rho^{*} -
      c/\rho^{*})+ \beta f^{ex}(\rho^{*}) +
      \ln \Lambda^{2}\rho^{*} - 1
    \end{equation}

    It should be noted that eq.(5) requires the knowledge of the
    free-energy at the reference density $\rho^{*}$. To calculate
    this, we use the Frenkel-Ladd~\cite{kn:ladd} method of coupling a solid to
    harmonic springs, referred to as an Einstein lattice.
    The benefit of coupling the solid in this way is that the free
    energy of the Einstein lattice can be easily calculated
    analytically, allowing one to obtain the free energy of the system
    of interest. We consider a system that is dependent upon a
    coupling parameter $\xi$ such that the total energy of the
    system may be written as

    \begin{equation}
        U(\xi) = U_{o} + \xi U = U_{o} + \xi \sum_{i = 1}^{N}(\vec{r}_{i} - \vec{r}_{o,i} )
    \end{equation}

    \noindent where N is the number of particles, $\vec{r}_{o,i}$
    the position of the lattice site to which the particle
    \emph{i} is assigned, $U_{o}$ the energy of the system of
    interest, and $\vec{r}_{i}$ are the positions of the particles.
    As the coupling parameter $\xi$ becomes large, the system
    becomes more strongly coupled to the lattice. For very large
    values the system will behave as a non-interacting Einstein
    lattice. To verify this behavior, the mean-squared
    displacements of the system for various values of $\xi$ are
    computed and compared to those of a non-interacting Einstein
    lattice. Once this value $\xi_{max}$ is obtained, we can
    calculate the free energy of the square well model at a
    reference density $\rho^{*}$ by~\cite{kn:excess}

    \begin{equation}
        F_{SW} = F(\xi_{max}) - \int_{0}^{\xi_{max}} d\xi <U(\vec{r}\,^{N},
        \xi)>_{\xi},
    \end{equation}

    \noindent where the first term represents the free energy of the
    Einstein lattice. With the chemical potentials of both phases now known, it
    is straightforward to calculate a coexistence point and
    to subsequently perform the Gibbs-Duhem method, thereby
    obtaining the fluid-solid coexistence curve.

    \begin{figure}
     \rotatebox{-90}{\scalebox{.5}{\includegraphics{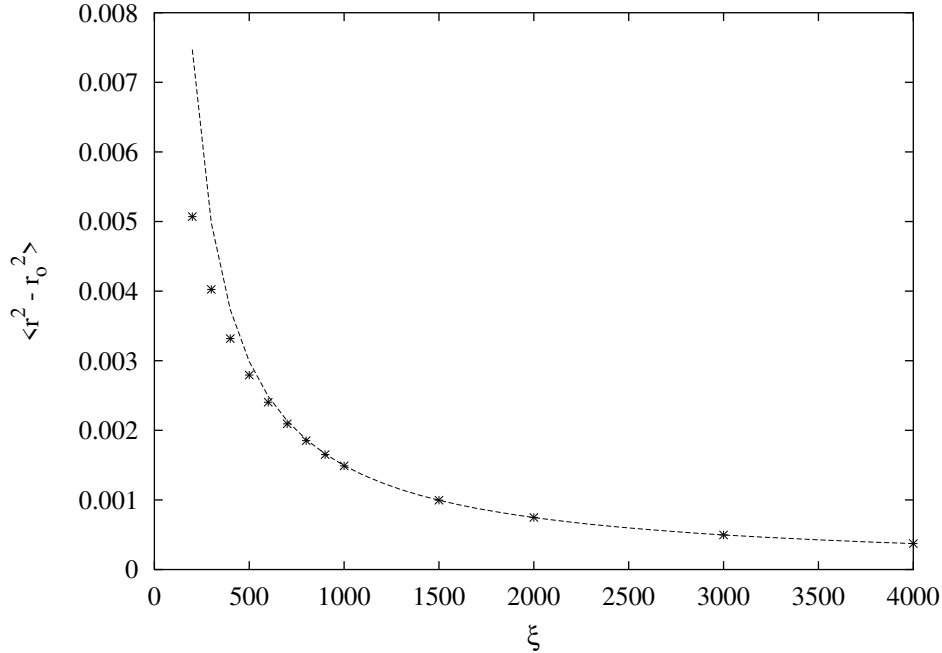}}}
     \caption{\label{fig:epsart}\small {Comparison of mean squared displacements for a coupled solid with
     that of an Einstein lattice (solid line) for the range $\lambda = 1.25$.}}
     \end{figure}

    To obtain a coexistence point, NPT simulations are first
    performed for $N = 256$ particles in a periodic simulation cell. In one NPT simulation, on average, one volume displacement is attempted
    for every N attempts at displacing a particle. This is done because a volume move is
    computationally more expensive than a particle displacement.
    Equilibration runs lasted for 10 million
    Monte Carlo steps and production runs ran for 20 million Monte Carlo steps, a step being an attempt at either displacing
    a particle or changing the volume of the cell, on average. In order to compare the chemical potentials for
    coexistence, we must first perform simulations on a system
    coupled to an Einstein lattice, as described previously. We
    perform regular Metropolis Monte Carlo simulations on a system
    of $N = 256$ particles with the constraint that the center of
    mass of the particles remains fixed~\cite{kn:frenkelbook}. If a
    particle is given a random displacement, all particles are
    subsequently shifted in the opposite direction to ensure that
    the center of mass is constant. We update the position of the
    center of mass every time a trial move is accepted. Thus, the
    shift in the center of mass is continually updated in order to
    properly calculate the harmonic energy contribution~\cite{kn:frenkelbook}.
    Einstein lattice simulations were equilibrated for 5 million
    Monte Carlo steps and were run for a total of 10 million Monte
    Carlo
    steps. These simulations are carried out at different values
    of the coupling parameter $\xi$. Fig. 1 shows our results for
    the mean squared displacements for $\lambda = 1.25$. We use $\xi_{max}
    = 4000$ at $\lambda = 1.25$ and $\xi_{max} = 3000$ at $\lambda = 1.15$ for the square well systems, respectively. The integral in
    eq. (5) is evaluated using a 10-point Gaussian quadrature~\cite{kn:recipe} to
    obtain the free energy of the square well system. Once this
    free energy is obtained, the chemical potentials are plotted versus pressure
    to obtain a coexistence point, given by
    the point at which the two curves intersect.

     \begin{figure}
     \rotatebox{-90}{\scalebox{.5}{\includegraphics{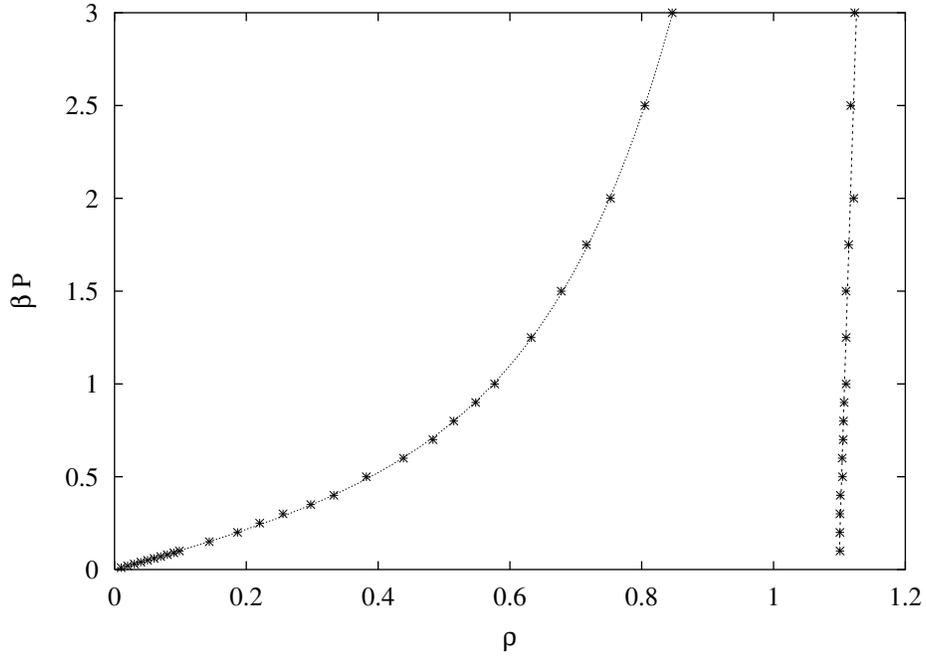}}}
     \caption{\label{fig:epsart}\small {Equation of state of the
     square well model for $\lambda = 1.15$. Also shown as the
     dashed lines are our fits according to equations in the
     text. Data was collected at the temperature T = 1.0.}}
     \end{figure}

     \begin{figure}
     \rotatebox{-90}{\scalebox{.5}{\includegraphics{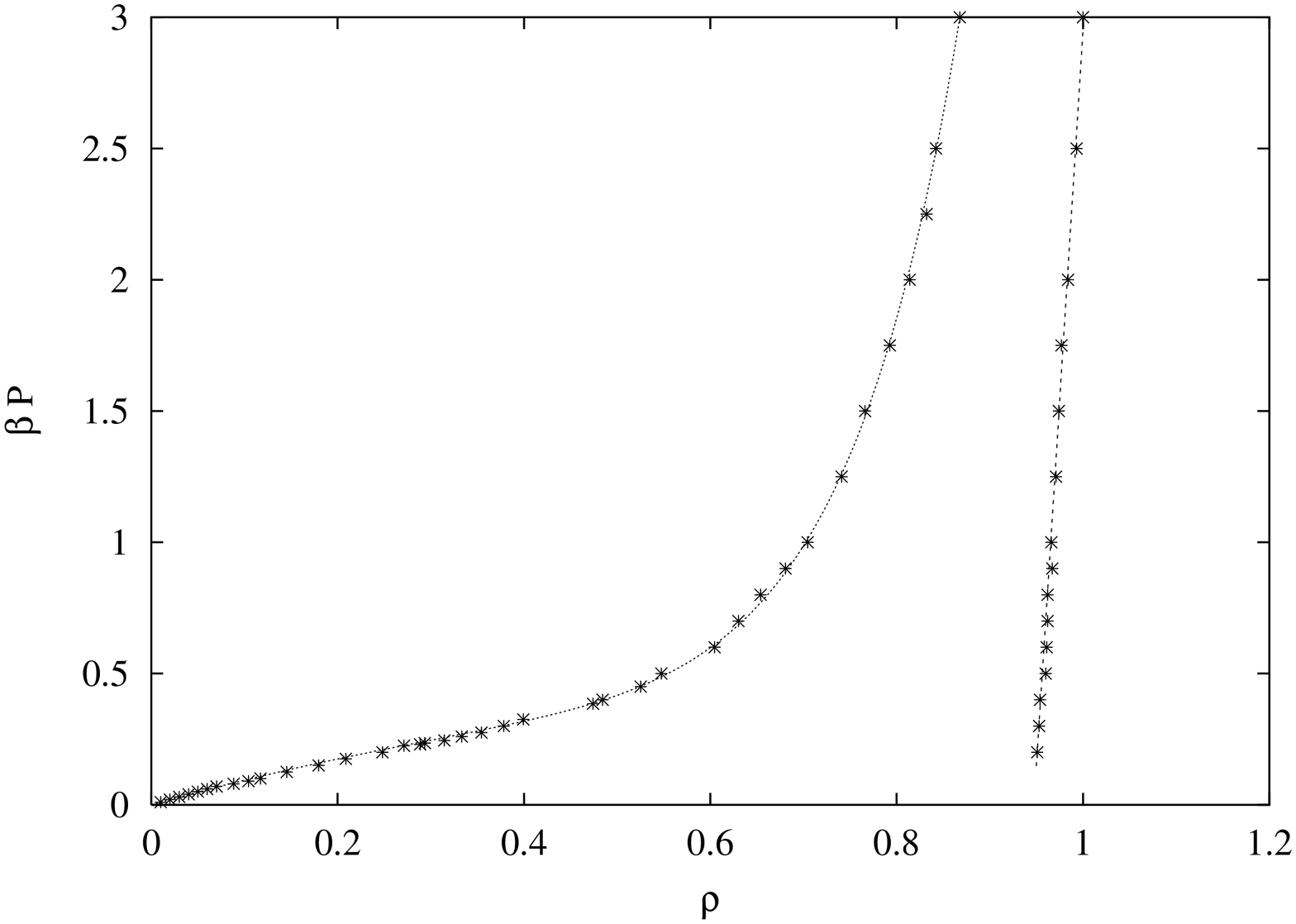}}}
     \caption{\label{fig:epsart}\small {Equation of state of the
     square well model for $\lambda = 1.25$. Also shown as the
     dashed lines are our fits according to equations in the
     text. Data was collected at the temperature T = 1.0.}}
     \end{figure}

     \begin{figure}
     \rotatebox{-90}{\scalebox{.5}{\includegraphics{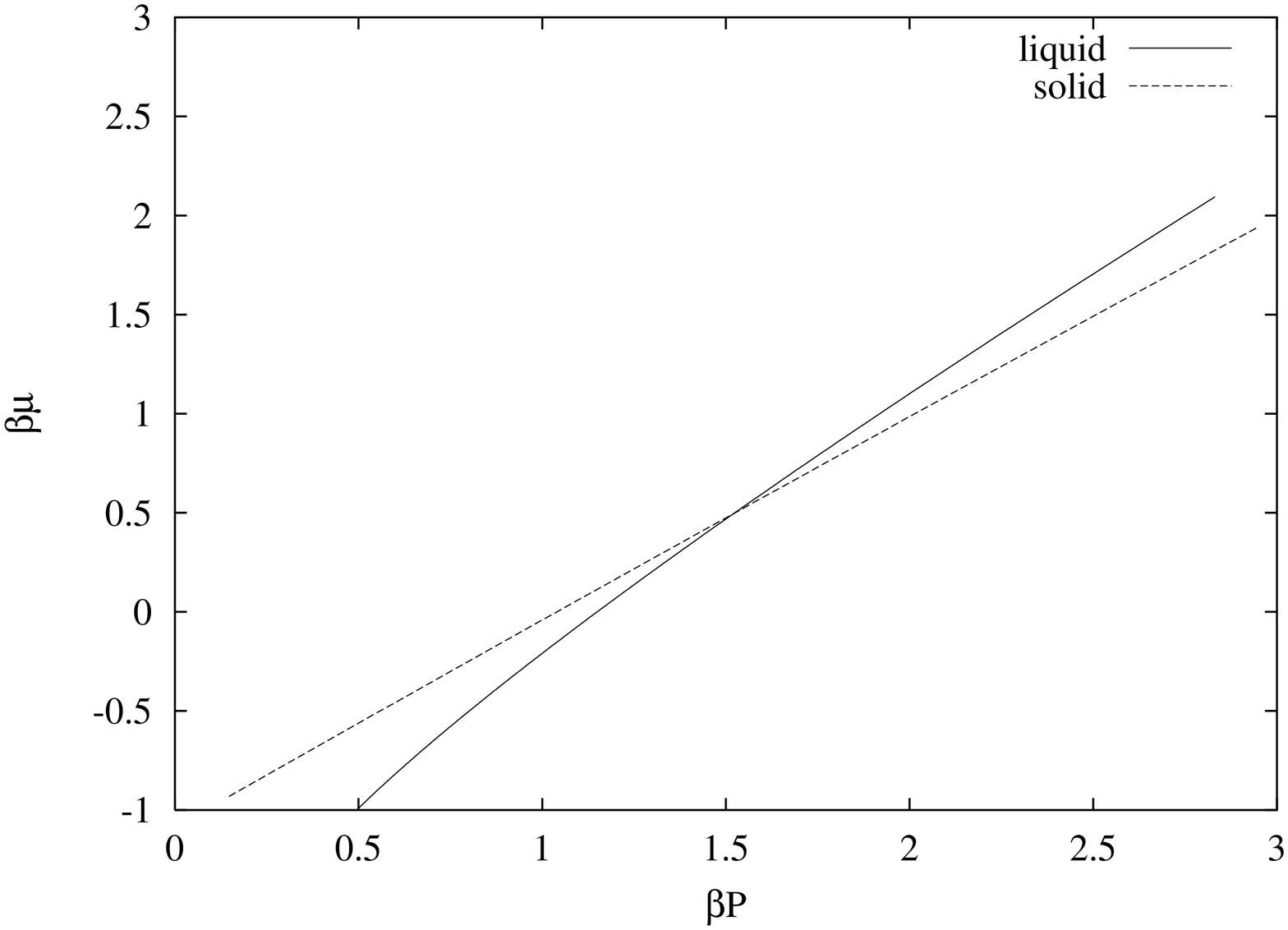}}}
     \caption{\label{fig:epsart}\small {Determination of a coexistence point for $\lambda = 1.25$. The point at which the curves cross satisfies
     the condition for coexistence, namely, equal chemical potentials at equal pressures.}}
     \end{figure}

        The Gibbs-Duhem integration is performed using two NPT
        simulations in parallel. Simulations were equilibrated for
        20 million Monte Carlo steps and ran for 40 million Monte Carlo steps in
        production. Each NPT simulation was performed under the
        same conditions as described previously. The coexistence
        pressures were calculated using a simple predictor-corrector
        algorithm~\cite{kn:recipe}. With the predicted/corrected pressure, NPT
        simulations are done in parallel to obtain the next
        coexistence point, and the process is repeated until the
        full fluid-solid coexistence curve is obtained. This
        procedure was implemented for both systems at $\lambda =
        1.15$ and $\lambda = 1.25$.

    \subsection {Fluid-fluid coexistence}

    To calculate fluid-fluid coexistence of the square well model
    at $\lambda = 1.25$, we use the Gibbs ensemble Monte Carlo
    method~\cite{kn:gibbs}. Two physically separated but thermodynamically
    connected simulation cells are used to calculate both the
    less dense and more dense fluid phases, respectively. This method circumvents
    the problem of an interface, which would hinder sampling, by
    altogether removing it. The two cells are allowed to exchange
    particles and both separately undergo volume displacements
    such that the total number of particles $N = N_{1} + N_{2}$ and
    total volume $V = V_{1} + V_{2}$ remain constant. Particles are also
    displaced within each cell according to the regular Metropolis
    method. Simulations for $N = 512$ particles were conducted on a
    periodic simulation cell. Simulations were equilibrated for 50
    million steps and produced for 100 million steps. We required
    that the chemical potentials of both phases at each state
    point be equal to ensure coexistence had been reached.

    At shorter interaction ranges, $\lambda$, it becomes increasingly
    difficult to sample phase space. The system becomes
    non-ergodic and standard techniques used to calculate fluid-fluid
    coexistence are unable to obtain coexistence points. As
    $\lambda$ is decreased, the critical temperature $T_{c}$ is
    decreased, and the particles tend to 'stick' together.
    Stickiness~\cite{kn:rosenbaum,kn:frenkelspheres} is a phenomenon usually associated with the
    adhesive hard-sphere model~\cite{kn:baxter}. It is not unusual, therefore, that
    for sufficiently short-range square-well potentials stickiness begins to manifest itself. (An adhesive sphere
    interaction can be derived from a square-well interaction in
    the limit where the well becomes infinitesimally narrow and infinitesimally
    deep.)

    To overcome this difficulty, we employ
    parallel-tempering~\cite{kn:geyer} Monte Carlo to speed up equilibration of the system.
    Parallel tempering allows local Monte Carlo simulations
    (replicas) to communicate and exchange information between each
    other. The benefit of this is that systems that were unable to
    properly sample phase space are able to do so. Parallel tempering is quickly becoming a standard
    method of sampling systems that become trapped in local energy minima~\cite{kn:yan,kn:frenkelspheres,kn:pagan}.
    We set up global Monte Carlo simulations using grand canonical Monte
    Carlo simulations in each replica, where each replica is
    allowed to exchange particle-configurations according to

    \begin{equation}
    p_{acc}(x_{i} \leftrightarrow x_{i+1}) = min[1, \exp(- \Delta \mu \Delta
    N)],
    \end{equation}

    \noindent where $x_{i}$ is the state of the $i^{th}$ replica and $\mu$ is the reduced chemical potential
    $\mu/k_{B}T$. The potential used is actually $\mu^{*} =\mu - 3\;ln[\Lambda/\sigma]$ (as in Ref~\cite{kn:tildesley}).
    In what follows, we denote $\mu^{*}$ as $\mu$ for simplicity.
    To ensure detailed balance is obeyed, a replica is selected at
    random and tested to swap particle-configurations with a
    neighboring replica according to the above probability.  Within
    each replica three trial moves are attempted: 1)
    insertion/deletion trial moves are attempted according to
    standard Monte Carlo as adapted for use in the grand canonical
    ensemble~\cite{kn:tildesley}, 2) particle displacements are employed according to
    the regular Metropolis method, and 3) cluster moves~\cite{kn:cluster} are
    attempted to break-up any small clusters that may have formed.
    The last trial move is similar to those of particle
    displacements, however, extra care must be taken so that
    detailed balance is observed. Thus, a cluster of particles is
    displaced such that no new clusters are formed and no old
    clusters are destroyed.

     \begin{figure}
     \rotatebox{-90}{\scalebox{.5}{\includegraphics{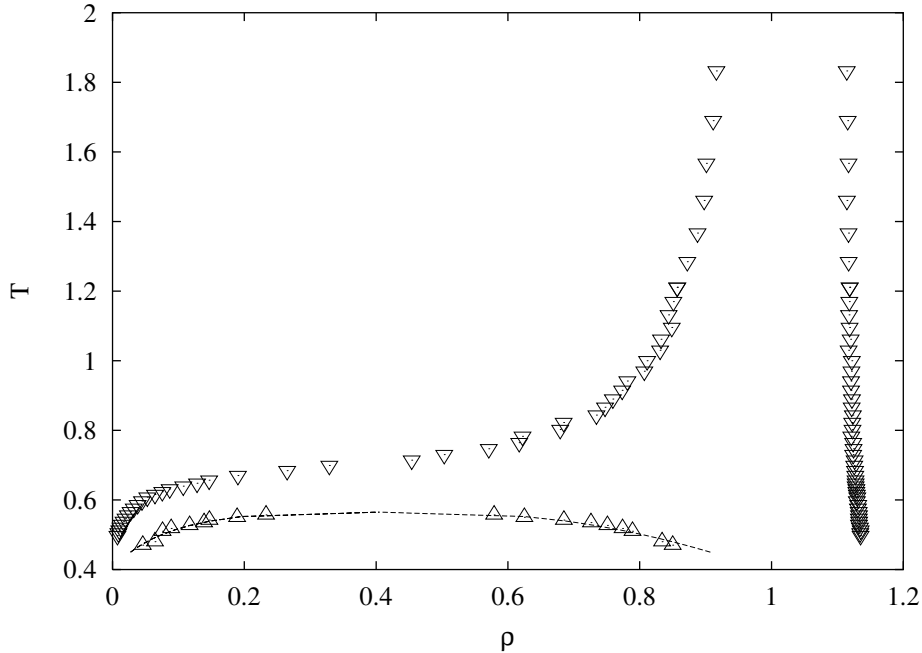}}}
     \caption{\label{fig:epsart}\small {Phase diagram of square well model for $\lambda = 1.15$ using Monte Carlo methods
     described in the text. Shown are the fluid-solid ($\triangledown$) and metastable fluid-fluid ($\triangle$) coexistence curves
     along with our fit (solid line) to the latter.}}
     \end{figure}

     \begin{figure}
     \rotatebox{-90}{\scalebox{.5}{\includegraphics{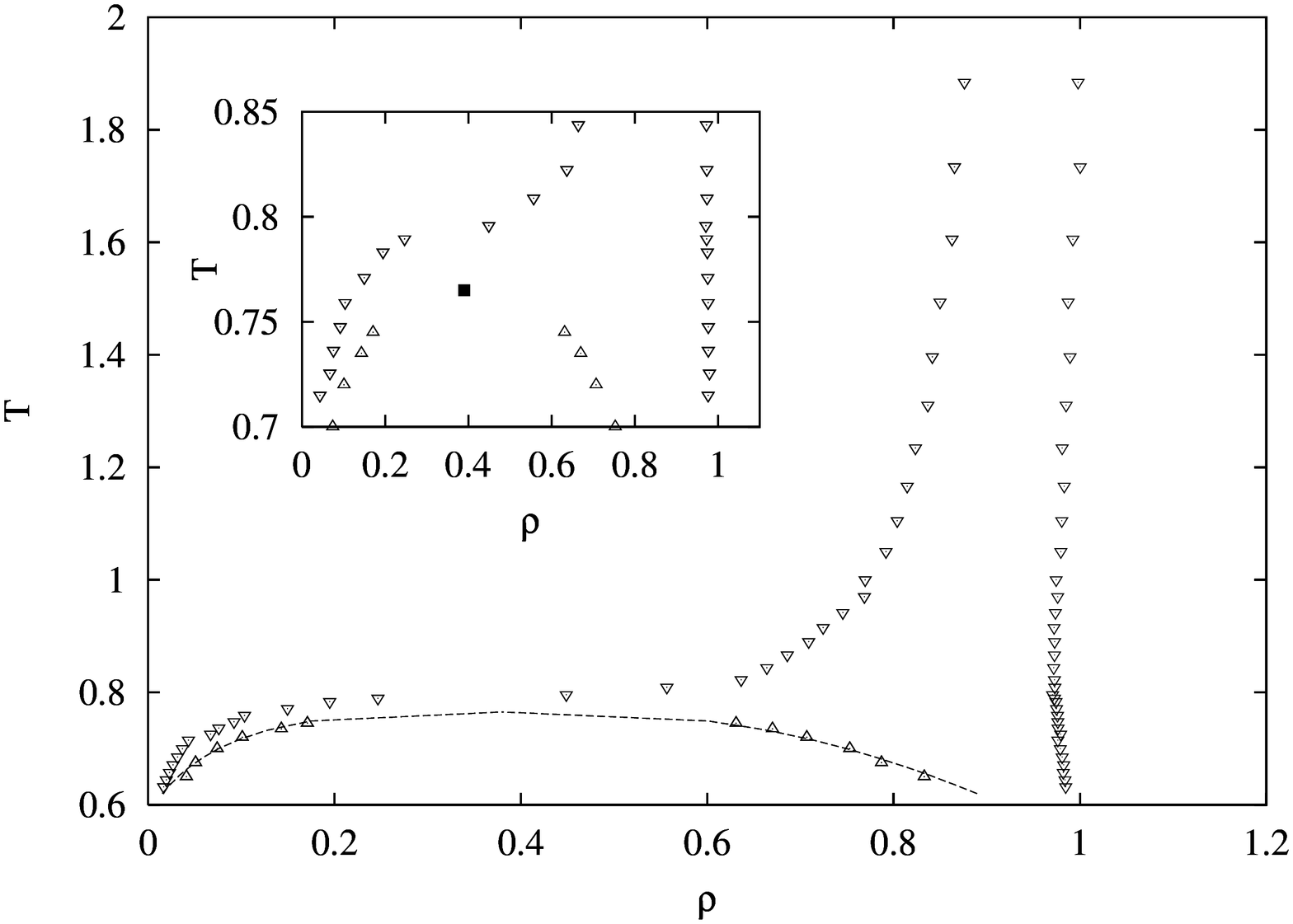}}}
     \caption{\label{fig:epsart}\small {Phase diagram of square well model for $\lambda = 1.25$ using Monte Carlo methods
     described in the text. Shown are the fluid-solid ($\triangledown$) and metastable
     fluid-fluid($\triangle$)
     coexistence curves along with our fit (solid line) to the latter. The inset shows a close-up view of the
     phase diagram near the estimated critical point ($\blacksquare$).}}
     \end{figure}

    Periodic boundary conditions are used for a simulation cell of
    size $L = 8\sigma$. Simulations were equilibrated for 10
    million steps and production runs lasted 100 million steps. At
    each state point we chose to run our replicas at a common
    temperature with each differing in chemical potential $\mu$.
    We used 6 to 7 replicas per global simulation, choosing the chemical
    potentials such that approximately $20\%$ of particle
    exchanges were accepted. Once a density distribution was
    obtained in this way, coexistence points were determined
    using an equal-area criterion according to

    \begin{equation}
    \int_{0}^{\langle\rho\rangle} P_{L}^{(\beta' \mu')}(\rho) = 0.5.
    \end{equation}

    To facilitate our grand canonical simulations, we employ
    histogram-reweighting~\cite{kn:histogram} after each set of simulations.
    This method allows information at one state point to be
    obtained from a neighboring one, such that

    \begin{eqnarray}
    &&P^{(\beta',\mu')}_{L}(\rho,u) = \nonumber \\
    && \frac{\exp[(\mu' - \mu)\rho V - (\beta' - \beta)u
    V]P^{(\beta, \mu)}_{L}(\rho,u)}{\sum \exp[(\mu' - \mu)\rho V - (\beta' - \beta)u
    V]P^{(\beta, \mu)}_{L}(\rho,u)}.
    \end{eqnarray}

    \section{Results and Discussion}

    \begin{table}
    \caption{\label{tab:table1}Results from fit to fluid-fluid coexistence curves}
    \begin{ruledtabular}
    \begin{tabular}{cccccc}
   $\lambda$ & $T_{c}$ & $\rho_{c}$ & $\phi_{c}$ & A & B\\
    \hline 1.15 & 0.565$(3)$ & 0.404$(3)$ & 0.212& 0.55$(2)$ & 0.89$(1)$\\
    1.25 & 0.765$(2)$ & 0.390$(4)$ & 0.204 & 0.50$(2)$ & 0.82$(1)$\\
    \end{tabular}
    \end{ruledtabular}
    \end{table}

    We calculated the phase diagrams of the square well model for
    both $\lambda = 1.15$ and $\lambda = 1.25$, respectively. Our equations of state for both interaction ranges,
    simulated at the isotherm $T = 1.0$, are shown in Figs. 2 and 3. Our fits to the data are also shown in these
    figures, as described previously in the text. We
    emphasize the importance of obtaining very good fits to the data, as one must have
    accurate values of the chemical potentials of the two phases, vis-a-vis eqs. (4) and
    (5). Errors in fitting the data, particularly for the liquid,
    lead to errors in the value for a coexistence point and,
    ultimately, in determining the fluid-solid coexistence phase
    boundaries. Care was taken such that deviations between the
    data and the fit were usually less than 1.0\%, no data point
    deviated from the fit by more than 3.0\% for either model. Certain data points were obtained by running longer
    simulations to minimize fluctuations.

    After obtaining good fits to our isotherms, we obtained
    coexistence
    points for the square well models. This is shown in Fig. 4 where a coexistence point for $\lambda = 1.25$
    has been determined by comparing the chemical potential versus pressure.
    We then employed the Gibbs-Duhem method for each model to obtain
    the fluid-solid coexistence curve. Figs. 5 and 6 show the
    phase diagrams for both interaction ranges studied. It can be
    seen that for $\lambda = 1.15$ the metastable fluid-fluid
    coexistence curve is far below the liquidus line. This curve
    clearly indicates that the threshold for metastability lies
    above above this value for the interaction range. Other studies~\cite{kn:stell,kn:bolhuis,kn:rascon} have been done on
the Yukawa and square well models
    for extremely short range interactions. There, a phenomenon of isostructural solid-solid coexistence is observed,
    with the appearance of a solid-solid critical point. We observe no such coexistence at $\lambda = 1.15$, although
    we have not extended our simulations to high densities where such a phenomenon has been observed at lower
    ranges.
    Fig. 6 shows
    the phase diagram for $\lambda = 1.25$, showing that the
    fluid-fluid coexistence curve is metastable. We note that
    at the liquidus line just above the critical temperature,
    $T_{c}$, simulations exhibited large fluctuations in the density. We therefore ran
    longer simulations in this region than for other parts of the
    fluid-solid phase diagram to minimize fluctuations.

     \begin{figure}
     \rotatebox{-90}{\scalebox{.5}{\includegraphics{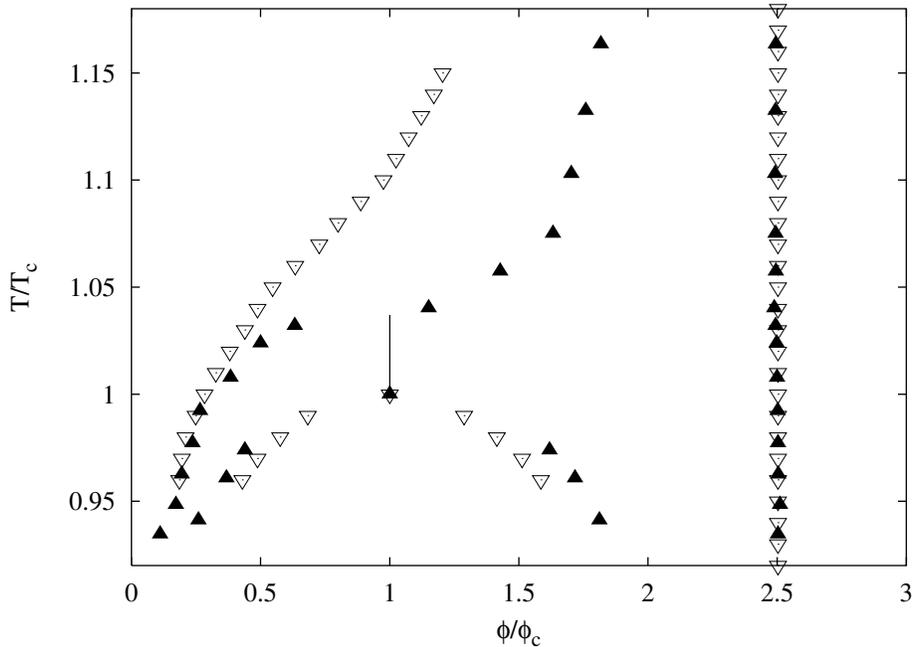}}}
     \caption{\label{fig:epsart}\small {Comparison of our Monte Carlo results ($\blacktriangle$) to the cell model results of Ref. ~\cite{kn:benedek} ($\triangledown$) for
     the range $\lambda = 1.25$.
     Also plotted are our estimates of the critical temperature $T/T_{c}$. The metastability gap is represented as a solid
     line.}}
     \end{figure}

    Also shown in Figs. 5 and 6 are our attempts to fit the metastable
    fluid-fluid coexistence curves to the equation $\rho_{\pm} - \rho_{c} = A|T - T_{c}| \pm
    B|T-T_{c}|^{\beta}$, where $T_{c}$ and $\rho_{c}$ are the
    critical temperature and density, respectively, and $\beta = 0.3258$ (Ref~\cite{kn:beta}) is
    the Ising expononet. Table I shows our values for both $\lambda =
    1.15$ and $\lambda = 1.25$. Our estimates compare well with
    other predictions~\cite{kn:benedek,kn:vega} of the critical point for these two
    ranges. The critical density $\rho_{c}$ is shifted toward a
    higher density for $\lambda = 1.15$ as compared to that at $\lambda = 1.25$, as shown in Table
    I, but is smaller than the critical density predicted in (Ref~\cite{kn:benedek,kn:MonteCarlo}).

    As noted in the introduction, many models capture the
    qualitative characteristics of protein phase diagrams. It is interesting, however,
    that the Yukawa, MLJ, and 2n-n Lennard Jones models all have
    been shown~\cite{kn:extended} to become metastable at the interaction range
    $R = \lambda -1 = 0.13 - 0.15$. As noted, this is not, however, the case
    for the square well model. Our results confirm that $R \simeq
    0.25$, as previous~\cite{kn:benedek,kn:daanoun} studies have predicted.

    \begin{figure}
     \rotatebox{-90}{\scalebox{.5}{\includegraphics{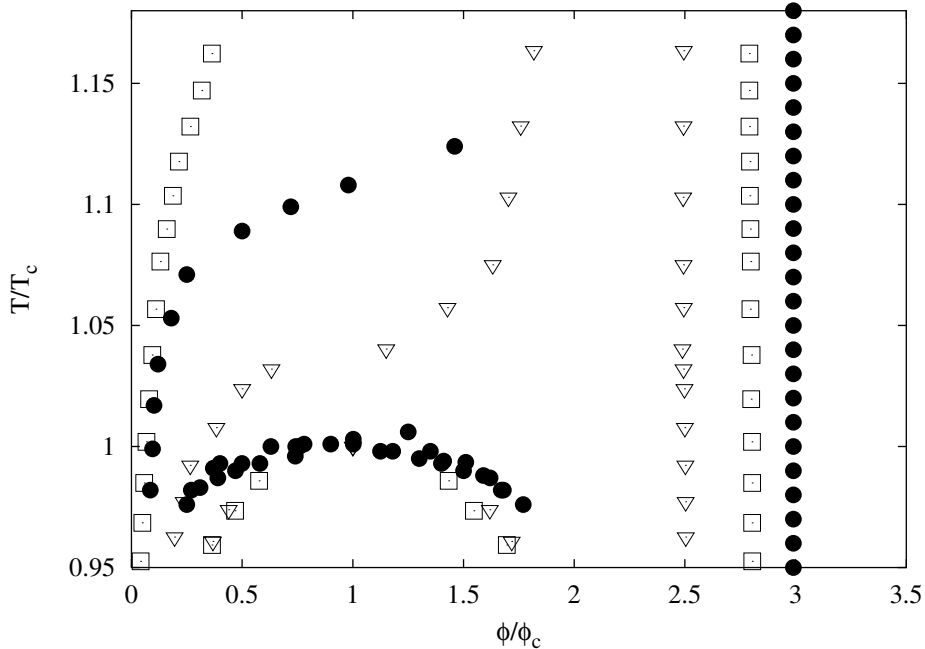}}}
     \caption{\label{fig:epsart}\small {Comparison of our Monte Carlo results for both $\lambda = 1.15$ ($\Box$) and
     $\lambda = 1.25$ ($\triangledown$), respectively, to the gamma-II crystallin ($\bullet$).}}
     \end{figure}

    In another study~\cite{kn:benedek} of the square well model at $\lambda = 1.25$,
    the authors obtain the phase diagram by a combination of Monte Carlo
    and extrapolation techniques~\cite{kn:MonteCarlo}. We compare our phase diagram at
    $\lambda = 1.25$ to theirs in Fig. 7, converting to the units
    $T/T_{c}$ and volume fraction $\phi/\phi_{c}$ ($\rho/\rho_{c}$).
    It is observed that both studies predict similar behavior for
    both the solidus line and metastable fluid-fluid coexistence
    curve. The liquidus lines, however, are not compatible.
    We quantify this discrepancy  by measuring the metastability
    gap, defined by~\cite{kn:benedek}

    \begin{equation}
    \frac{T_{L} - T_{c}}{T_{c}} = \frac{n_{s} - n_{s}^{*}}{n_{s}^{*} - n_{l}}
    \end{equation}

    \noindent where $n_{i}$ are phenomenological values used in their model.
    Using this definition we find the gap values 0.240 for $\lambda =
    1.15$ and .035 for $\lambda = 1.25$ for the metastability gaps. The former predictions~\cite{kn:benedek} overestimate
    the metastability gap by approximately 50\% and 60\%,
    respectively. The fact that our results for the solidus line
    agree with the early study is probably due to the fact that
    their choice for the number of contacts a given particle has
    within its interaction range in their cell model approximation is consistent with the fcc
    crystal structure for this model.

    Much interest has been focused on the gamma crystallins~\cite{kn:gamma}, a
    group of monomeric eye lens proteins. Certain mutations~\cite{kn:mutation} of
    these proteins has been linked to genetic cataracts and protein crystallization~\cite{kn:mutation2}. These
    proteins have been well studied~\cite{kn:crystallins} and their phase behavior are
    similar to those of the square well model, warranting a comparison
    of the two. Fig. 8 shows our square well phase diagrams to
    those~\cite{kn:benedek} of the gamma-II crystallins. Although we compare our data with the gamma-II crystallin, it has been
shown~\cite{kn:MonteCarlo} that the data for the fluid-fluid
coexistence curve of the entire gamma crystallin family lies on
the same curve, within the scatter of the data. Our values for
$\phi_c$, shown in Table I, are close to that of the value for the
family of gamma crystallins. Both models qualitatively capture the
characteristics of the gamma protein,
    with the
    model at the range $\lambda = 1.15$ more closely resembling it. The case $\lambda = 1.15$ yields better agreement
 with the liquidus and solidus lines than the model with $\lambda = 1.25$. For both models, however, the fluid-fluid coexistence
curve is not as broad as that of the gamma-II crystallin.
 Indeed, some studies~\cite{kn:aeolotopic,kn:kern} have attempted to remedy this by including anisotropy in the
    interactions among particles in the square well model, thereby broadening the fluid-fluid coexistence
    curves. One study~\cite{kn:aeolotopic} averages out the rotational degrees of freedom to obtain
    an effective temperature-dependent isotropic potential that better
    approximates both the liquidus and metastable fluid-fluid coexistence curves of the
    gamma-IIIb crystallin.

    \section{Conclusion}

    The square well model for the interaction ranges $\lambda =
    1.15$ and $\lambda = 1.25$ have been obtained using Monte
    Carlo simulations, including parallel tempering. We find that the
    latter interaction range is just below the threshold value for
    metastability of the fluid-fluid coexistence curve.
    Finite-size effects have not been taken into
    account for either the liquidus line or metastable fluid-fluid
    coexistence curve near the critical point where critical fluctuations
    could affect our results. However, the metastability gap is
    sufficiently large that we believe our conclusion about the
    metastable nature of the coexistence curve for $\lambda = 1.25$
    is correct.
    We have also compared our Monte Carlo results for
    the complete phase diagram at $\lambda = 1.25$ to another study~\cite{kn:benedek} and find that this study does not
predict the correct liquidus line. The
    complete phase diagram at $\lambda = 1.15$ was obtained by
    using standard Monte Carlo techniques, overcoming the problem
    of 'stickiness' by using parallel tempering for the
    fluid-fluid coexistence curve. A comparison of these two
    phase diagrams with the gamma-II crystallin has been made.
    Better qualitative agreement is obtained at $\lambda = 1.15$ with the experimental results
for the gamma-II crystallin, but there is no quantitative agreement.

    Isotropic fluids cannot be expected
    to quantitatively match experimental systems. Although certain characteristics of the phase
    diagrams of proteins can be captured by these simple models, it is unlikely that a correct solidus line
    can be obtained. This is due to anisotropic interactions and the (non spherical) ellipsoidal shape of certain globular proteins,
    such as the gamma crystallins~\cite{kn:gammaShape}. Models such as the square well
    model, or similar ones, alone, cannot account
    for effects such as anisotropy, solvent-solute interactions, or
    hydration effects commonly found in experimental systems and are, thus, too simple to be
    quantitatively accurate. However, the isotropic models do seem
    to provide a useful first approximation which can then be
    extended to include effects such as anisotropy and
    hydrophobicity.

    \section{Acknowledgements}

    This work is supported by NSF grant DMR0302598.


\begin{thebibliography}{99}
\bibitem{kn:gast} A. P. Gast, C. K. Hall, and W. B. Russel, J.
Colloid Interface Sci., \textbf{96}, 251 (1983)

\bibitem{kn:leker}H. N. W. Lekkerkerker \emph{et al}., Europhys. Lett.,
\textbf{20}, 559 (1992); Tejero et al., Phys. Rev. Lett.,
\textbf{73}, 752 (1994)

\bibitem{kn:experiment1} S. M. Ilett \emph{et al}., Phys. Rev. E, \textbf{51}, 1344
(1995)

\bibitem{kn:experiment2} M. Muschol and F. Rosenberger, J. Chem. Phys., \textbf{107},
1953 (1997)

\bibitem{kn:rosenbaum}D. Rosenbaum, P.C. Zamora, and C.F. Zukoski,
Phys. Rev. Lett., \textbf{76}, 150 (1996)

\bibitem{kn:yukawa} M. H. J. Hagen and D. Frenkel, J. Chem. Phys.
\textbf{101}, 4093, (1994)

\bibitem{kn:tenwolde} P. R. ten Wolde and D. Frenkel, Science
\textbf{277}, 1975 (1997)

\bibitem{kn:oxtoby} D. W. Oxtoby and V. Talanquer, J. Chem. Phys.
\textbf{101}, 223 (1998)

\bibitem{kn:benedek} N. Asherie, A. Lomakin, and G. B. Benedek, Phys. Rev. Lett.
\textbf{77}, 4832 (1996)

\bibitem{kn:extended} Massimo G. Noro and Daan Frenkel, J. Chem.
Phys. \textbf{113}, 2941 (2000)

\bibitem{kn:vega}L. Vega, E. de Miguel, L. F. Rull, G. Jackson and I. A. McLure, J. Chem. Phys.\textbf{96}, 2296 (1992)

\bibitem{kn:daanoun} A. Daanoun, C.F. Tejero, and M. Baus, Phys.
Rev. E \textbf{50}, 2913 (1994); C. Rasc\'{o}n, G. Navasqu\'{e}s,
and L. Mederos, Phys. Rev. B \textbf{51}, 14899 (1995)

\bibitem{kn:crystallins} M. L. Broide \emph{et al}., Proc. Natl.
Acad. Sci. USA \textbf{88}, 5660 (1991); C. R. Berland \emph{et
al}., \emph{ibid}, \textbf{89}, 1214 (1992)

\bibitem{kn:mutation2} S. Kmoch \emph{et al.}, Human Molecular Genetics \textbf{9},
1779 (2000); A. Pande \emph{et al.}, Proc. Natl. Acad. Sci. USA
\textbf{98}, 6116 (2001)


\bibitem{kn:dixit} N. M. Dixit and Charles F. Zukoski,
Journal of Colloid and Interface Science \textbf{228}, 359 (2000)

\bibitem{kn:stell} G. Foffi \emph{et al}., Phys. Rev. E
\textbf{65}, 031407 (2002)

\bibitem{kn:kofke} D. A. Kofke, J. Chem. Phys. \textbf{98},
4149 (1993); \emph{ibid}, Mol. Phys. \textbf{78}, 1331 (1993)

\bibitem{kn:note} We use different forms of the Clausius-Clapeyron
equation at high and low temperatures to minimize numerical errors
as in Refs. (~\cite{kn:kofke,kn:noro,kn:yukawa})

\bibitem{kn:noro} M. G. Noro and D. Frenkel, J. Chem.
Phys. \textbf{114}, 2477 (2001)

\bibitem{kn:ladd} D. Frenkel and A.J.C. Ladd, J. Chem. Phys.
\textbf{81}, 3188 (1984)

\bibitem{kn:excess} To correct for finite-size effects for
crystals with a fixed center of mass, we calculate the free-energy
contribution for discontinuous potentials as in the following
reference: J. M. Polson \emph{et al}., J. Chem. Phys.
\textbf{112}, 5339 (2000)

\bibitem{kn:frenkelbook} D. Frenkel and B. Smit,\emph{Understanding Molecular
Simulation}, (San Diego, Academic Press 2002)

\bibitem{kn:recipe} Numerical Recipes Online, World Wide Web,
\emph{http://www.library.cornell.edu/nr}

\bibitem{kn:gibbs} A. Z. Panagiotopoulos, Mol. Phys. \textbf{61},
813 (1987)

\bibitem{kn:baxter} R. J. Baxter, J.Chem. Phys.
\textbf{49}, 2770 (1968)

\bibitem{kn:frenkelspheres} M. A. Miller and D. Frenkel, Phys.
Rev. Lett. \textbf{90}, 135702, (2003)

\bibitem{kn:geyer} C. J. Geyer and E. A. Thompson, J. Am. Stat.
Assoc. \textbf{90}, 909 (1995)

\bibitem{kn:tildesley} M. P. Allen and D. J. Tildesley,
\emph{Computer Simulations of Liquids}, (Clarendon, Oxford, 1990)\

\bibitem{kn:yan} Q. Yan and J. J. de Pablo, J. Chem. Phys.
\textbf{111}, 9509 (1999)

\bibitem{kn:pagan} D. L. Pagan, M. E. Gracheva, and J. D. Gunton,
J. Chem. Phys. \textbf{120}, 8292 (2004)

\bibitem{kn:cluster} N. A. Seaton and E. D. Glandt, J. Chem.
Phys. \textbf{84}, 4595 (1986)

\bibitem{kn:histogram} R. H. Swendsen, Physica A \textbf{194},
53, (1993)

\bibitem{kn:bolhuis} P. and D. Frenkel, Phys. Rev.
Lett. \textbf{72}, 2211 (1994)

\bibitem{kn:rascon} C. Rasc\'{o}n, L. Mederos, and G. Navascu\'{e}s, Phys.
Rev. Lett. \textbf{77}, 2249 (1996); \emph{ibid}, J. Chem. Phys.
\textbf{103}, 9796 (1995)

\bibitem{kn:beta}A. M. Ferrenberg and D. P. Landau, Phys. Rev. B \textbf{44},
5081 (1991)

\bibitem{kn:gamma} J. Graw, Biol. Chem. \textbf{378}, 1331 (1997)

\bibitem{kn:mutation} D. A. Stephan \emph{et al.}, Proc.
Natl. Acad. Sci. USA \textbf{96}, 1008 (1999); A. Pande \emph{et
al.}, Proc. Natl. Acad. Sci. USA \textbf{97}, 1993 (2000)


\bibitem{kn:aeolotopic} A. Lomakin, N. Asherie, and G.
B. Benedek, Proc. Natl. Acad. Sci. USA, \textbf{96}, 9465 (1999)

\bibitem{kn:kern} N. Kern and D. Frenkel, J. Chem. Phys.
\textbf{118}, 9882 (2003)

\bibitem{kn:gammaShape} Ajit Basak \emph{et al.}, J. Mol. Biol.
\textbf{328}, 1137 (2003)

\bibitem{kn:MonteCarlo} A. Lomakin, N. Asherie, and G. B. Benedek,
J. Chem. Phys. \textbf{104}, 1646 (1996)





\end{thebibliography}
\end{document}